\documentclass[11pt,a4paper]{article}
\usepackage[T1]{fontenc}
\usepackage[utf8]{inputenc}
\usepackage{lmodern,amsmath,amssymb,microtype}
\usepackage[margin=25mm]{geometry}
\usepackage{authblk}
\usepackage{xcolor}
\usepackage{xurl}
\usepackage[colorlinks=true,allcolors=blue]{hyperref}
\NewDocumentCommand\citeproctext{}{}

\makeatletter
\let\@cite@ofmt\@firstofone
\def\@biblabel#1{}
\def\@cite#1#2{{#1\if@tempswa , #2\fi}}
\makeatother
\newlength{\cslhangindent}
\newlength{\csllabelwidth}
\newenvironment{CSLReferences}[2]
 {\begin{list}{}{\setlength{\itemindent}{0pt}\setlength{\leftmargin}{0pt}
  \setlength{\parsep}{0pt}
  \ifodd #1\setlength{\leftmargin}{\cslhangindent}\setlength{\itemindent}{-1\cslhangindent}\fi
  \setlength{\itemsep}{#2\baselineskip}}}
 {\end{list}}
\usepackage{calc}

\title{FLITS: an interactive, reproducible workbench for fast radio
burst analysis}
\author[1]{Dirk Kuiper}
\author[1, 2]{Ziggy Pleunis}
\author[1, 2, 3, 4]{Jason W. T. Hessels}
\affil[1]{Anton Pannekoek Institute for Astronomy, University of
Amsterdam, Science Park 904, NL-1098 XH Amsterdam, the Netherlands}
\affil[2]{ASTRON, Netherlands Institute for Radio Astronomy, Oude
Hoogeveensedijk 4, 7991 PD Dwingeloo, the Netherlands}
\affil[3]{Department of Physics, McGill University, 3600 University
Street, Montreal, Quebec H3A 2T8, Canada}
\affil[4]{Trottier Space Institute, McGill University, 3550 University
Street, Montreal, Quebec H3A 2A7, Canada}
\date{26 October 2026}
\hypersetup{pdftitle={FLITS: an interactive, reproducible workbench for
fast radio burst
analysis},pdfauthor={Dirk Kuiper, Ziggy Pleunis, Jason W. T. Hessels}}
\begin{document}
\maketitle
\section{Summary}\label{summary}

Fast radio bursts (FRBs) are brief flashes of radio emission, usually
detected from distant galaxies (Lorimer et al., 2007; Petroff et al.,
2022). They are commonly studied using a dynamic spectrum: a record of
the radio signal as a function of time and observing frequency.
Measurements of burst brightness, duration, frequency structure, and
polarization help researchers investigate the source and the material
through which its radiation has travelled.

FLITS (Fast-Look Interactive Transient Suite) is an open-source Python
package with a browser interface for analysing detected bursts. It
brings data inspection, interference removal, measurement, and modelling
into a single interactive session. It reads the data, masks
interference, and fits the \texttt{fitburst} (Fonseca et al., 2024)
burst model itself, so users can apply each step to the same selected
data without moving between separate programs. Other analysis packages
can work on the same prepared data, which is available from Python and
in exported files.

FLITS saves the measurement settings, selected data regions, excluded
frequency channels, and calibration inputs in a portable session file.
Together with the original data, this file allows users to reopen an
analysis and recompute supported measurements. Outputs describe how
uncertainties were estimated and flag missing calibration information,
making published results easier to check and reproduce.

\section{Statement of need}\label{statement-of-need}

For each detected burst, radio observations provide dynamic spectra,
often for the four Stokes parameters: total intensity (\(I\)), linear
polarization (\(Q\) and \(U\)), and circular polarization (\(V\)). Some
observations retain only total intensity. Turning these data into an
astrophysical interpretation requires measurements such as dispersion
measure (DM), which describes the delay of lower-frequency radiation by
free electrons, and rotation measure (RM), which describes the
wavelength-dependent rotation of linear polarization by magnetized
plasma. Other quantities include burst width, arrival time, spectrum,
fluence (flux density integrated over time), and, given calibration and
distance information, peak luminosity. Scintillation bandwidth and
scattering time characterize fine frequency structure and pulse
broadening caused by propagation through intervening material.

These derived properties are routinely reported in papers, but a table
of values rarely records every choice needed to reproduce them. Results
depend on the burst and background regions, excluded interference, time
and frequency resolution, calibration, and fitted model. Comparing
observations from instruments with different formats and sensitivities
adds further complexity. The campaign that motivated FLITS combines data
from the Green Bank Telescope at L and P band, Nançay, Westerbork,
Onsala, Stockert, and CHIME.

Automated pipelines are essential for samples of hundreds to thousands
of bursts, but unusual or particularly informative events often warrant
individual attention. For example, successive components in some
repeating FRBs appear at progressively lower frequencies, the
``sad-trombone'' effect (Hessels et al., 2019). This structure can be
confused with an incorrect dispersion correction, so the DM that
maximizes signal-to-noise need not best preserve the burst's components.
Interactive inspection lets researchers compare DM choices, separate
overlapping components, select uncontaminated background data, and check
what a fitted model fails to explain. It supports careful assessment of
such ambiguities without claiming to remove them automatically.

FLITS combines this flexibility with a record of the settings used.
Users can refine an individual analysis, compare measurements across
observations, and share the saved session and original data so
collaborators can examine how the results were obtained.

\section{State of the field}\label{state-of-the-field}

Several packages address the analysis of bursts after detection.
\texttt{fitburst} (Fonseca et al., 2024) models dynamic spectra to
estimate burst shape, dispersion, and scattering parameters. FRBGui
(Chamma et al., 2023) provides interactive measurements of burst
structure in time and frequency. DM\_phase (Seymour et al., 2019)
estimates DM using resolved burst structure, while RM-Tools (Van Eck et
al., 2026) provides rotation-measure synthesis and polarization fitting.
Stingray (Bachetti et al., 2024) offers broader time-series and
variability analysis, including power spectra, correlations, and
statistical modelling of astronomical light curves. These tools provide
complementary ways to study transient signals.

FLITS brings data preparation and its own implementations of several of
these methods together in a shared session. Its readers are tested
against \texttt{your} (Aggarwal et al., 2020), its automatic masking
follows the per-channel statistical tests of \texttt{jess} (Kania et
al., 2026), and its burst-model fits are tested against
\texttt{fitburst}. Its own RM-synthesis implementation is checked
against an RM-Tools (Van Eck et al., 2026) reference dataset. A separate
workbench is useful because these methods are otherwise spread across
several specialist packages and file formats, each preparing the data
separately. New file formats are added as reader plugins without changes
to FLITS itself, and other tools can work on the session's prepared data
through the Python interface or exported files. Its main contribution is
to keep an analysis's selections and settings alongside its
measurements, so the interactive workflow can be inspected and supported
calculations repeated.

\section{Software design}\label{software-design}

The same Python analysis routines serve both the browser interface and
command-line workflows. Changes made in the browser update a session
containing the data and analysis settings; the calculations can also run
without opening a browser. This allows automated tests to exercise the
measurement routines directly.

For larger samples, \texttt{flits\ measure} applies the same steps to a
list of files without a browser: it masks interference, locates the
burst in time and frequency, and measures signal-to-noise, peak flux
density, fluence, widths, and arrival times. Each result records a
status and warning flags, so bursts whose automatic selections need
checking can be reopened in the interface. \texttt{flits\ demo}
downloads a reference burst and opens it in the browser.

FLITS reads SIGPROC filterbank files and PSRFITS search and folded data
(Hotan et al., 2004), and supports CHIME/FRB HDF5 products, including
public catalogue dynamic spectra (CHIME/FRB Collaboration, 2021) and
beamformed power data. These readers convert the input into common
arrays and descriptive information. Additional file formats can be
supported through reader plugins.

A saved JSON session file records the input-file location and SHA-256
checksum, DM, data crop, burst and background regions, frequency range,
excluded channels, calibration inputs, and available results. This takes
more care to maintain than a table of measurements, but preserves the
settings needed to revisit an analysis. On reopening, FLITS checks the
source data against the recorded checksum and metadata. The file
captures the saved state, not a history of every interaction;
reproducing the calculation also requires the original data and an
appropriate software environment.

The \texttt{flits\ replay} command reopens the data and recomputes the
core measurements and any recorded width comparisons, frequency-drift
measurements, and polarization analyses performed on the session data.
It can compare the core numerical results with their saved values within
a specified tolerance. The saved drift-analysis settings include the
Monte Carlo random seed. DM scans, other time- and frequency-structure
analyses, and optional model fits remain saved results rather than being
rerun by this command.

Exports use JSON for structured results, CSV for tables, NumPy NPZ
archives for numeric and string arrays, PNG/SVG for plots, and optional
SIGPROC files for selected data windows. They do not serialize Python
objects with pickle. These formats allow stored results to be inspected
independently of FLITS, but do not guarantee identical recalculation
after software changes. Snapshots and exports automatically record
FLITS, Python, and installed dependency versions, with source revisions
where available. Each analysis retains its originating software record
when a session is reopened or other measurements are recomputed; missing
provenance in older files remains explicitly unknown. Tests use files
generated by earlier releases to check preservation of stored results
and detect changes during recalculation. Reproducing an analysis still
requires preserving its software environment: version records identify
software but do not archive it. The repository supplies pinned
dependencies and a public reference workflow.

Trial DMs can be scored by signal-to-noise, by the burst-structure
criterion of DM\_phase (Seymour et al., 2019), or by the
smoothed-structure metric of Sutinjo et al. (2023), which also reports
the range of DMs that the noise cannot distinguish from the best one.

Polarization analysis accepts suitable full-Stokes polarization input
files with known conventions, or separately prepared \(Q/U\) spectra.
FLITS performs weighted RM synthesis (Brentjens \& Bruyn, 2005; Heald,
2009), which combines the polarization across frequency to estimate
Faraday rotation. Instrumental polarization calibration and corrections
for Earth's ionosphere must be applied separately. Flux and fluence
uncertainties distinguish measurement noise from calibration
uncertainty. When uncertainty in the system-equivalent flux density
(SEFD), a measure of telescope sensitivity, is not supplied, the outputs
flag the incomplete uncertainty budget. These checks help users judge a
result's limits; they do not establish its scientific validity on their
own.

User documentation and reproducible examples are available at
\url{https://dirkkuiper.github.io/flits/}.

\section{Research impact statement}\label{research-impact-statement}

FLITS made the burst measurements for a study of the brightest bursts of
the repeating FRB 20240114A observed with several telescopes (Kuiper et
al., in prep.), and J. Huang uses it in a separate study of the same
source (Huang et al., in prep.). It is also used in a pulsar study of
subpulse morphology and DM optimization (Kuiper et al., in prep.) and by
A. Moroianu to analyse how the DM and burst morphology of precisely
localized repeating FRBs evolve over time (Moroianu et al., in prep.).
C. Makarytchev and V. Shah use it to determine accurate DMs for bursts
from FRB 20240209A.

Researchers outside the development team have contributed code. M. A.
Fine added support for folded PSRFITS pulsar data, treating pulse phase
as the time axis so that pulse profiles can be modelled. J. Huang
corrected how FLITS reads polarization data from the Nançay Radio
Telescope.

The \href{https://dirkkuiper.github.io/flits/guided-workflow/}{public
GBT reference workflow} provides a downloadable burst cutout, a saved
session, expected measurements, and pinned dependencies. Automated
checks exercise both the published package and the current source
checkout against this reference.

\section{AI usage disclosure}\label{ai-usage-disclosure}

Generative AI tools (OpenAI Codex/ChatGPT, Anthropic Claude Code, Google
Gemini, and Microsoft Copilot) assisted with software development,
documentation, and manuscript preparation, including prose editing and
review. Specific model versions were not systematically recorded;
however, use generally involved a then-current flagship model from the
respective provider. The authors reviewed and validated AI-assisted
outputs, made the core scientific and design decisions, and retain
responsibility for the work.

\section{Acknowledgements}\label{acknowledgements}

FLITS builds on the scientific Python ecosystem, particularly NumPy
(Harris et al., 2020), SciPy (Virtanen et al., 2020), Astropy (Astropy
Collaboration, 2022), and Matplotlib (Hunter, 2007). Its readers and
interference masking replace \texttt{your} (Aggarwal et al., 2020) and
\texttt{jess} (Kania et al., 2026), which earlier versions used.

The AstroFlash research group at McGill University, University of
Amsterdam, ASTRON, and JIVE is supported by: a Canada Excellence
Research Chair in Transient Astrophysics (CERC-2022-00009); an Advanced
Grant from the European Research Council (ERC) under the European
Union's Horizon 2020 research and innovation programme (`EuroFlash';
Grant agreement No.~101098079); an NWO-Vici grant (`AstroFlash';
VI.C.192.045); an NSERC Discovery Grant (RGPIN-2025-06681); an ERC
Starting Grant (`EnviroFlash'; Grant agreement No.~101223057); and an
NWO-Veni grant (VI.Veni.222.295).

The funders had no role in the software design, implementation,
validation, or preparation of this manuscript.

\section*{References}\label{references}
\addcontentsline{toc}{section}{References}

\protect\phantomsection\label{refs}

\end{document}